\documentclass[lettersize,journal]{IEEEtran}
\usepackage{amsmath,amsfonts}
\usepackage{algorithmic}
\usepackage{algorithm}
\usepackage{array}
\usepackage[caption=false,font=normalsize,labelfont=sf,textfont=sf]{subfig}
\usepackage{textcomp}
\usepackage{stfloats}
\usepackage{url}
\usepackage{verbatim}
\usepackage{graphicx}
\usepackage{cite}
\begin{document}

\title{\fontsize{24}{28}\selectfont Dichroic Dual-Angle Refractor: Multi-Cell Huygens' Metasurface-Based  Circuit Approximation}


\author{
	\IEEEauthorblockN{Georgios Kyriakou\IEEEauthorrefmark{1}, Giampaolo Pisano\IEEEauthorrefmark{1}}
	
\IEEEauthorblockA{\IEEEauthorrefmark{1} Department of Physics, 
	University of Rome `La Sapienza', Rome, Italy \\
	\{georgios.kyriakou,giampaolo.pisano\}@uniroma1.it}}



\pagenumbering{gobble}

\maketitle

\begin{abstract}
A dichroic dual-angle refractor based on a 2D Huygens' metasurface model is treated by means of resonant-circuit approximations of multiple cells. The whole parameter space of refraction angles is tested for the analytic approximations involved, whereas the realised refraction guides the choice of optimal such angles. Our results indicate the possibility of sythesising a rectangular array of a number of cells which eventually depends on the low-frequency band diffraction limit. Aberrations from the ideal performance are also examined and attributed to certain approximation caveats.

\end{abstract}


\section{Introduction}
\IEEEPARstart{M}{etamaterial} technology is nowadays a mature field of research, and many modern systems employ components with engineered, naturally abnormal properties stemming from metamaterial-based designs. Huygens' metasurfaces, in particular, are based in generalised sheet transition conditions (GSTC) \cite{epstein2016} and provide an appealing surface impedance and admittance description of the characteristics of a metamaterial thin sheet, which ensure the field discontinuity via Huygens' meta-atoms. Circuit equivalent modelling techniques can then match the field-averaged quantities to tangible network representations \cite{selvanaygam2013}.

In radio-astronomical receivers, front-end RF chains usually need to deal with multiple optical paths to accomodate for dual-band operation, which in turn proves challenging and prone to wave diffraction anomalies whenever these paths have to be synthesised using reflectors. A dichroic dual-angle refractor such as the one envisioned in our work would deflect the incoming beam in different directions using only one thin sheet, avoiding previous complex designs and ensuring minimum losses due to reflection. The design of such a device follows similar techniques established for astronomical instrumentation \cite{pisano2015}.

In this work, we report advances in the design of a fully-transmissive surface achieving plane-wave refraction at two diverse angles in two different frequency bands, first presented in \cite{kyriakou2025}. The unit-cell scattering parameters across a finite array are numerically simulated with respect to their transmission and reflection properties and their realised refraction angle, which is then compared to the design angles chosen in the two bands. Exploring the parameter space of possible design angles, we identify the optimal ones while highlighting the limitations of our approach. 

\section{Design Updates and multi-cell array S-parameter approximation results}

It is well-known in metasurface literature that multi-band behaviour of a subwavelength-cell array is primarily achieved in the presence of multiple resonances of the constituent unit-cells \cite{xu2018,ginis2018}. In our work, two resonant circuits, a series LC (electrically resonant) and a parallel LC (magnetically resonant) are collocated to achieve the desired dual-band performance. Symmetry is retained by using the X-equivalent circuit with identical port characteristic impedances as that of the TE-polarised wave refraction \cite{wong2016}.

As has been investigated in \cite{kyriakou2025}, approximating the homogenised representation of the ideal unit-cell impedance matrix of the dichroic dual-angle refractor only at its first Laurent series term defines all the circuit parameters and presents tolerable errors. This is sufficient and avoids approximation of further terms that limit the parameter space by constraining the two refraction angles. In this regard, these angles \( \theta_{t,l}\) and \( \theta_{t,h} \) are free to choose and uniquely define the two circuits, contrary also to stricter implementations such as in \cite{dorrah2018}.

As an example of the approximation of the scattering parameters of a multi-cell array, employing the formulas of \cite{kyriakou2025}, we assume a distribution of unit-cells on the \( x\)-axis in the interval \( [2.2~{\rm mm}, 7.4~{\rm mm}]\) with a \( 0.4 \)~mm step. It also proved beneficial for our modelling to now associate both \( \omega_e,\ \omega_m\) poles to trigonometric function poles \( \omega_s\neq0\) of the homogenised expressions, in order to shift the low-band to higher frequencies and avoid approaching the diffraction limit. This meant multiplying \cite[Eq.~(7)]{kyriakou2025} by 1/2 (to resemble Eq.~(11)) and redefining the electic resonance as \( \omega_e=s(x)\pi c_0/(x\sin\theta_{t,l})\)\footnote{\( s(x)=[x] \) is used throughout (rounding operation)}. 

\begin{figure}
    \centering
    \subfloat[\( |S_{11}^{\rm uc}|\)]{\includegraphics[width=\linewidth]{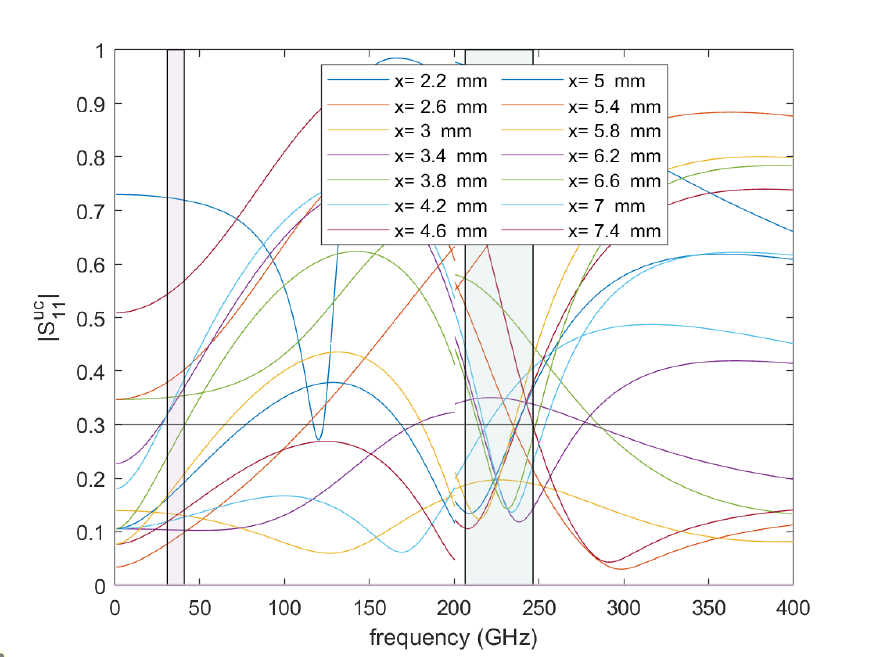}}
    \quad
    \subfloat[\( |S_{12}^{\rm uc}|\)]{\includegraphics[width=\linewidth]{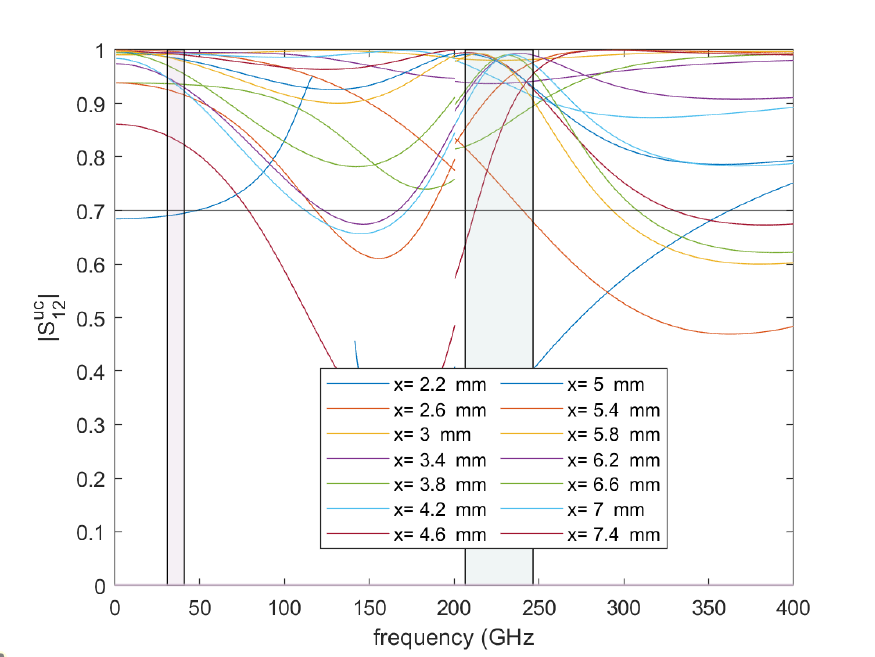}}
    \quad
    \subfloat[\( \theta_t^{\rm uc} \) (deg)]{\includegraphics[width=\linewidth]{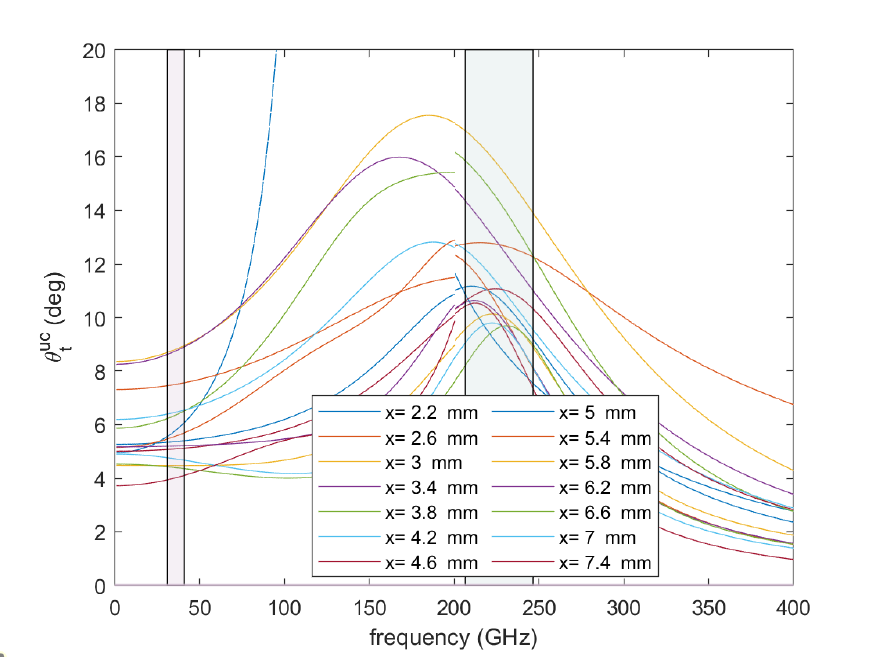}}
    \caption{\( |S_{11}^{\rm uc}|,\ |S_{11}^{\rm uc}|\) and \( \theta_{t}^{\rm uc}\) across frequency for \( 2.2~{\rm mm}\leq x \leq7.4~{\rm mm}\) with a 0.4~mm step. Highlighted are tentative bounds for the low-band and high-band operation. }
    \label{fig:sparameters}
\end{figure}

Fig.~\ref{fig:sparameters} presents the unit-cell S-parameters for each longitudinal distance, as constructed using the circuit approximation for the respective Z parameters with the design angles \( \theta_{t,l}=76^\circ,\ \theta_{t,h}=63^\circ \) identified as optimal. It is important to note that in transforming from \( Z\) to \( S\) parameters, the port characteristic impedance \( Z_0 \) continuously changes value capturing the spatial dispersion phenomena arising from the approximation. We arbitrarily choose the center of the full frequency range (200~MHz) as the transformation `jump' from  \( Z_0=\eta_0/\cos(\theta_{t,l}) \) to \( Z_0=\eta_0/\cos(\theta_{t,h}) \). The realised refraction angle is then calculated as:
\begin{equation}
    \theta_{t}^{\rm uc}={\rm asin}\left(\frac{c_0}{x}\frac{{\rm d(Arg}(S_{12}^{\rm uc}){\rm )}}{{\rm d}\omega}\right)
\end{equation}

 We have first confirmed that \( S_{11}^{\rm uc}=S_{22}^{\rm uc}\) as expected from the formulation used. We then observe that the reflection remains mostly at levels below -10~dB (0.3) for most of the cell positions, with deviations due to the limited bandwidth of the resonance and the rounding error at non-integer positions. Similarly, the transmission \( S_{12}^{\rm uc}\) remains close to unity for most of the cells. 

 What is most important to note though is that the realised refraction angle in the third plot varies significantly within the frequency range and does never coincide with the design angle for each band. This discrepancy is again a symptom of the approximation strategy and the spatial dispersion it introduces. Finally, we calculate these realised angles as statistical means \( \mu_x(\theta_{t,l})=5.65^\circ,\ \mu_x(\theta_{t,h})=11.21^\circ \) across \( x \), with respective stds \( \sigma_x(\theta_{t,l})=1.4^\circ,\ \sigma_x(\theta_{t,h})=1.94^\circ \). This means that two distinct angles are still effectively seen in the two bands.

\section{Conclusions}
We have presented recent numerical modelling results of a circuit-approximated array of meta-atoms of a dichroic dual-angle refractor. Our multi-cell approximation is prone to inaccuracies for inner meta-atom positions and generally the design angles are overestimations of the realised angles, but those are still distinguishable from each other in the identified low and high bands. This ongoing work will be complemented by electromagnetic simulations of the proposed array of cells during the conference presentation, considering further practical implementation aspects.

\section{Acknowledgment}
This work has been financially supported by Italian gonvernment funds from PRIN 2022 (Progetti di Ricerca di Rilevante Interesse Nazionale) grant no.~2022AYFS7F.

\vspace{3ex}

\vfill


\vspace{-1em}
\begin{thebibliography}{1}
\bibliographystyle{IEEEtran}
\bibitem{epstein2016}
A.~Epstein and G.~Eleftheriades, ``Huygens' metasurfaces via the equivalence principle: design and applications," in \emph{J. Opt. Soc. Am. B}, vol.~33, A31-A50 (2016)

\bibitem{selvanaygam2013}
M.~Selvanayagam and G.~V.~Eleftheriades, ``Circuit Modeling of Huygens Surfaces," in \emph{IEEE Antennas and Wireless Propagation Letters}, vol. 12, pp. 1642-1645, 2013, doi: 10.1109/LAWP.2013.2293631

\bibitem{pisano2015}
G.~Pisano, C.~Tucker, P.~A.~R. Ade, P.~Moseley and M.~W.~Ng, ``Metal mesh based metamaterials for millimetre wave and THz astronomy applications," \emph{2015 8th UK, Europe, China Millimeter Waves and THz Technology Workshop (UCMMT)}, Cardiff, UK, 2015, pp. 1-4, doi: 10.1109/UCMMT.2015.7460631

\bibitem{kyriakou2025}
G.~Kyriakou, G.~Pisano, L.~Olmi and F.~Piacentini, ``Design of a Dichroic Transmissive Huygens' Metasurface Unit-Cell Presenting Refraction Angle Duality," in \emph{19th European Conference on Antennas and Propagation (EuCAP)}, Stockholm, Sweden, 2025, pp~ 1-5, doi: 10.23919/EuCAP63536.2025.10999362

\bibitem{xu2018}
G.~Xu, S.~V.~Hum and G.~V.~Eleftheriades, ``A Technique for Designing Multilayer Multistopband Frequency Selective Surfaces," in \emph{IEEE Transactions on Antennas and Propagation}, vol. 66, no. 2, pp. 780-789, Feb. 2018, doi: 10.1109/TAP.2017.2772089.

\bibitem{ginis2018}
V.~Ginis, P.~Tassin, T.~Koschny, C.~M.~Soukoulis, ``Broadband metasurfaces enabling arbitrarily large delay-bandwidth products" in \emph{Applied Physics Letters}, vol. 108, no. 3 (2016), doi:10.1063/1.4939979

\bibitem{wong2016}
J.~P.~S.~Wong, A.~Epstein and G.~V.~Eleftheriades, ``Reflectionless Wide-Angle Refracting Metasurfaces," in \emph{IEEE Antennas and Wireless Propagation Letters}, vol. 15, pp. 1293-1296, 2016, doi: 10.1109/LAWP.2015.2505629

\bibitem{dorrah2018}
A.~H.~Dorrah and G.~V.~Eleftheriades, ``All-pass characteristics of a Huygens' unit cell," in \emph{2018 United States National Committee of URSI National Radio Science Meeting (USNC-URSI NRSM)}, Boulder, CO, USA, 2018, pp. 1-2.



\end{thebibliography}
\end{document}